\documentstyle[12pt]{article}

\newcommand{\lbd}{\lambda}
\newcommand{\vp}{\varphi}
\newcommand{\be}{\begin{equation}}
\newcommand{\ee}{\end{equation}}
\newcommand{\om}{\omega}

\begin{document}

\begin{center}
{\Large{\bf Excited Coherent Modes of Ultracold Trapped Atoms} \\ [5mm]
V.I. Yukalov$^{1,2}$, E.P. Yukalova$^{1,3}$, and V.S. Bagnato$^1$} \\ [3mm]

{\it $^1$Instituto de Fisica de S\~ao Carlos, Universidade de S\~ao Paulo \\
Caixa Postal 369, S\~ao Carlos, S\~ao Paulo 13560--970, Brazil \\ [3mm]
$^2$Bogolubov Laboratory of Theoretical Physics \\
Joint Institute for Nuclear Research, Dubna 141980, Russia \\ [3mm]
$^3$Laboratory of Computing Techniques and Automation  \\
Joint Institute for Nuclear Research, Dubna 141980, Russia}

\end{center}

\vskip 2cm

\begin{abstract}

A method of exciting coherent spatial modes of Bose--condensed trapped
atoms is considered. The method is based on the resonance modulation of the
trapping potential. The population dynamics of coherent modes is analysed.
The method makes it possible to create mixtures of different spatial modes
in arbitrary proportions, including the formation of pure excited coherent
modes. Novel critical effects in the population dynamics are found.

\end{abstract}

\newpage

\section{Introduction}

Bose atoms in magnetic traps can be cooled down to ultralow temperatures
where Bose--Einstein condensation takes place. There exists numerous
literature, both theoretical and experimental, devoted to this subject (see
reviews [1,2]).

The Bose condensation occurs when the thermal wavelength of atoms,
$\lbd\equiv\sqrt{2\pi\hbar^2/m_0k_BT}$, becomes larger than
the mean interparticle distance, $a$. At the same time, the effective radius
of an atom, $a_0$, has to be much smaller than the interatomic distance in
order that strong hard--core repulsion would not disturb much the motion
of colliding particles. When the thermal wavelength is such that
$$
\lbd\;\gg \; a \; \gg a_0 \; ,
$$
atoms are mutually correlated and a coherent state develops.

In the system under equilibrium, condensed atoms are in the ground state.
Recently [3], the idea was advanced of the possibility to create
non--ground--state condensates corresponding to excited coherent modes
of trapped atoms. This can be done by imposing, in addition to the trapping
potential, a time--dependent field whose oscillation frequency is adjusted
to be in resonance with the transition frequency between the ground--state
level and a chosen excited level. In this report, we present the results
of investigation of the population dynamics in a coherent system of trapped
atoms, driven by a resonance field. As it turned out, the dynamics of such
a system is quite nontrivial and is rather different from the population
dynamics of two--level optical systems. Thus, unusual effects, that can be
called critical, are found when a small variation of system parameters
results in a drastic change of the population dynamics.

\section{Population Dynamics}

Consider the Bose gas of neutral atoms at ultralow temperatures, when all
atoms are in a condensed state. The wave function of coherent atoms is
obtained from the nonlinear Schr\"odinger equation
\be
i\hbar\; \frac{\partial\vp}{\partial t} = \left [ \hat H(\vp) +
V_{res}\right ] \vp \; ,
\ee
in which $\hat H(\vp)$ is a nonlinear Hamiltonian and  $V_{res}$ is a
resonance field. The gas is assumed to be diluted, and the interatomic
interactions are shape independent and can be modelled by the contact
potential
$$
\Phi(\vec r\;)= \;\frac{4\pi\hbar^2a_s}{m_0}\;\delta(\vec r\;) \; ,
$$
where $m_0$ is atomic mass and $a_s$ is the scattering length.
Then the nonlinear Hamiltonian is
\be
\hat H(\vp) = \; - \; \frac{\hbar^2}{2m_0}\; {\vec\nabla}^2 +U(\vec r\;)
+ A|\vp|^2 \; ,
\ee
where $U$ is a trapping potential and
$$
A\equiv \; 4\pi\hbar^2 \; \frac{a_s}{m_0}\; N \; ,
$$
represents the effective interaction for $N$ confined particles.
The resonance field can be taken in the form
\be
V_{res} = V(\vec r\;) \cos\omega t \; .
\ee

Let us note that the nonlinear Schr\"odinger equation is an {\it exact}
equation of coherent states [4]. Contrary to this, if $\vp$ is treated as
the order parameter associated with the condensate, then Eq. (1) is an
{\it approximate} equation corresponding to the mean--field approach at zero
temperature. Such an approximate equation is often called the
Gross--Ginzburg--Pitaevskii equation [5--7].

The stationary solutions of Eq. (1), if the time--dependent field is absent,
are $\vp_n\exp(-iE_nt/\hbar)$ with $\vp_n$ and $E_n$ defined by the
eigenvalue problem
\be
\hat H(\vp_n)\vp_n = E_n\vp_n \; .
\ee
The eigenfunctions $\vp_n$, labelled by a multi--index $n$, are stationary
{\it coherent modes}. The transition frequencies $\om_{mn}$ between two
energy levels are given by the difference
\be
\hbar\omega_{mn} \equiv E_m - E_n \; .
\ee

Assuming that at the initial time $t=0$, all atoms are in the ground--state
coherent mode, we have the initial condition
\be
\vp(\vec r,0) = \vp_0(\vec r\;) \; .
\ee
And let then the resonant field (3) be switched on, with a frequency $\omega$
being in resonance, or almost in resonance, with the transition frequency
\be
\om_{j0} \equiv\; \frac{E_j - E_0}{\hbar}
\ee
between the ground state and a chosen energy level $j$. The corresponding
quasiresonance condition is
\be
\left | \frac{\Delta\omega}{\omega_{j0}}\right | \ll 1\; , \qquad
\Delta\omega\equiv\om -\om_{j0} \; .
\ee

The solution of the time--dependent equation (1) can be presented as an
expansion
\be
\vp(\vec r,t) = \sum_n\; c_n(t)\vp(\vec r\;)\; \exp\left ( -
\frac{i}{\hbar}\; E_n t\right )
\ee
over the coherent modes $\vp_n(\vec r\;)$. Combining Eqs. (1) and (9), we
find a system of equations for the coefficients $c_n(t)$. This system can be
simplified in the resonance approximation giving
$$
\frac{dc_0}{dt} = \; - i\; \alpha n_j c_0 -\; \frac{i}{2}\; \beta
e^{i\Delta\om t}\; c_j \; ,
$$
\be
\frac{dc_j}{dt} = \; - i\; \alpha n_0 c_j -\; \frac{i}{2}\; \beta^*
e^{- i\Delta\om t}\; c_0 \; ,
\ee
where
\be
n_i(t) \equiv |c_i(t)|^2
\ee
is the population of the level $i$, and a new notation is introduced for the
interaction amplitude
\be
\alpha\equiv \; \frac{A}{\hbar}\; \int\; |\vp_0(\vec r\;)|^2
|\vp_j(\vec r\;)|^2 \; d\vec r
\ee
as well as for the transition amplitude
\be
\beta \equiv \; \frac{1}{\hbar}\; \; \int\; \vp_0^*(\vec r\;) \;
V(\vec r\;)\; \vp_j(\vec r\;)\; d\vec r \; .
\ee
Because at the initial time all atoms are in the ground state, the initial
conditions to Eqs. (10) are
\be
c_0(0) = 1\; , \qquad c_j(0) = 0 \; .
\ee
Since the quantities $c_0$ and $c_j$ are complex, Eqs. (10) are to be
complemented by the equations for the complex conjugate $c_0^*$ and
$c_j^*$ or by the equations for the populations (11). The latter equations
are
$$
\frac{dn_0}{dt} = \; {\rm Im} \left ( \beta e^{i\Delta\om t}\; c_0^*c_j
\right ) \; ,
$$
\be
\frac{dn_j}{dt} = \; {\rm Im}\left ( \beta^* e^{-i\Delta\om t}\;
c_j^* c_0\right ) \; ,
\ee
with the initial conditions
\be
n_0(0) = 1 \; , \qquad n_j(0) = 1 \; .
\ee
In addition, the normalization condition
\be
n_0(t) + n_j(t) = 1
\ee
holds true.
The derivation of Eqs. (10)--(17) was expounded in detail in Ref. [3].

The system of nonlinear differential equations (10) and (15) can be solved 
analytically by means of the averaging technique [8] yielding for the
populations
\be
n_0 = 1 - \; \frac{|\beta|^2}{\Omega^2}\; \sin^2\frac{\Omega t}{2} \; ,
\qquad n_j = \; \frac{|\beta|^2}{\Omega^2}\; \sin^2\frac{\Omega t}{2} \; ,
\ee
in which $\Omega$ is an effective frequency given by the equation
\be
\Omega^2 =\left [ \alpha(n_0 - n_j) -\Delta\om \right ]^2 + |\beta|^2 \; .
\ee
The quantity $\Omega$ is an equivalent for the Rabi frequency, although
one has to keep in mind that it is actually not a fixed frequency but a
function of time defined by Eqs. (18) and (19). The solutions (18), for
$|\beta|<\alpha$, give a qualitative picture of complicated nonlinear
oscillations of the populations (11). In the limit $|\beta|\ll\alpha$, these
solutions become asymptotically exact.

Equations (15) show that if in some moment of time the pumping resonance
field (3) is switched off, then after this the populations $n_0$ and $n_j$
remain constant at their instantaneous values. This suggests the way of
creating mixtures of coherent modes with different spatial configurations,
and even pure excited coherent modes. Such excited coherent modes will not,
of course, last for ever. Their lifetime is defined by the total lost rate
$\gamma_j$ caused by atomic collisions. The same, actually, concerns the
ground--state condensate whose lifetime is defined by the corresponding
lost rate $\gamma_0$. The lost rates due to binary collisions can be
presented as
$$
\gamma_j =\lbd_{jj}N^2n_j^2\; \int\; |\vp_j(\vec r\;)|^4\; d\vec r +
\lbd_{j0}N^2n_jn_0\; \int\; |\vp_j(\vec r\;)|^2|\vp_0(\vec r\;)|^2 \;
d\vec r \; ,
$$
\be
\gamma_0 =\lbd_{00}N^2n_0^2\; \int\; |\vp_0(\vec r\;)|^4\; d\vec r +
\lbd_{0j}N^2n_0n_j\; \int\; |\vp_0(\vec r\;)|^2|\vp_j(\vec r\;)|^2 \;
d\vec r \; ,
\ee
where $\lbd_{ij}$ are the corresponding relaxation coefficients. The lost
rates due to ternary collisions can be presented in the similar way.
The ternary loss rates are usually much smaller than the binary ones [9].

The approximate solutions (18) and (19), obtained by the averaging method,
give us a general understanding of the excitation procedure. However, Eqs.
(18) and (19) define the level populations not explicitly but through a
connected system of equations. In order to study the dynamics of the
populations (11) explicitly, and also not to be limited by the
averaging--technique approximation, we return back to the evolution
equations (10), which we solve numerically. For numerical analysis, it is
convenient to measure time in units of $\alpha^{-1}$ and to introduce
dimensionless quantities
$$
b\equiv \; \frac{|\beta|}{\alpha} \; , \qquad \delta\equiv\;
\frac{\Delta\om}{\alpha} \; .
$$
The results of numerical calculations for several values of parameters are
presented in Figs. 1 to 6.

It turned out, that the dynamics of the populations is rather nontrivial
and exhibits a kind of critical effects when an anomalous coherent
oscillation of the state populations suddenly appears. The latter occur
on the critical line defined approximately as $b_c =0.5-\delta$. Outside
the critical line, the populations oscillate according to the sine--squared
law, as in Fig.1. Approaching the critical line, for instance by changing
the detuning, the period of oscillations suddenly increases by a jump, with
the top of $n_j$ and, respectively, the bottom of $n_0$, becoming flat. Thus,
the slight change of the detuning, from $\delta=0.11$ in Fig.1 to
$\delta=0.11017478$ in Fig.2, results in an abrupt period doubling. And
the following small shift of the detuning to $\delta=0.11017479$ results
again in the period being approximately doubled, with the qualitative change
of the time--dependence: Each second downward cusp between two adjacent
oscillations of $n_j$ overturns up becoming an upward cusp, which yields to
an effective period doubling, as is seen from comparing Figs. 2 and 3. The
further increase of the detuning drives the system away from the critical
line resulting in the decrease of the oscillation period and in the
oscillation behaviour again more resembling the sine--squared shape, as in
Figs. 4 and 5. The overall picture is qualitatively the same, when we cross
the critical line at other values of the parameters $b$ and $\delta$.
Thus, Fig.6 shows the time behaviour of the populations at $b=0.499$ and
$\delta=0.001001002$ being on the critical line. Figure 6 is similar to
Fig. 2, except that the oscillation amplitudes are different, being equal to
different values of $b$.

In practice, the value of $b$, which is the dimensionless expression for
the transition amplitude (13), depends on the resonant field (3). The
spatial part of the latter can be chosen to have different forms. For
example, in the consideration of a mixture of two Bose condensates of
$^{87}$Rb atoms in two internal hyperfine states, en effective potential
forcing the excitation of a first antisymmetric mode was linear [10],
due to the spatial separation of the condensate components. In our scheme,
the potential $V(\vec r\;)$ can be arbitrary.

\section{Conclusion}

We presented the analysis of temporal behaviour of spatial coherent modes
excited by a resonant field realizing the oscillatory modulation of the
trapping potential. Such a procedure of using the resonant pumping field
suggests the way of creating mixtures of coherent spatial modes in arbitrary
proportions, including the formation of pure excited coherent modes. 

There exists the
critical line connecting the values of the transition amplitude and detuning
at which the dynamics of fractional populations suddenly changes. The
qualitative and sharp change in the population dynamics reminds critical
phenomena occurring in equilibrium systems. It is possible to construct an
effective stationary system describing the averaged behaviour of populations
satisfying the original evolution equations. The effective averaged system
displays critical behaviour at the critical line which corresponds to that
observed for the nonequilibrium system. Effective critical indices can also
be defined. The detailed study of the critical dynamics is presently under
investigation and will be published elsewhere. The possibility of exciting
various coherent modes of Bose atoms can be important in the context of
realizing different spatial modes of atom lasers [11--15].

\vskip 5mm

{\bf Acknowledgement}

\vskip 2mm

We are grateful to financial support from the S\~ao Paulo State Research
Foundation (Fapesp) and the program Pronex.

\newpage

\begin{center}
{\large{\bf Figure captions}}
\end{center}

\vskip 5mm

{\bf Fig.1.} The populations of the excited coherent mode (solid line) and
of the ground--state mode (dashed line) as functions of time for $b=0.4$
and $\delta=0.11$.

\vskip 5mm

{\bf Fig.2.} The critical dynamics of the populations for $b=0.4$
and $\delta=0.11017478$.

\vskip 5mm

{\bf Fig.3.} Overturning of the downward cusps of $n_j$ upward, with the
period of oscillations being approximately doubled, occuring at $b=0.4$
and $\delta=0.11017479$.

\vskip 5mm

{\bf Fig.4.} The time dependence of the populations outside the critical
line, with $b=0.4$ and $\delta=0.12$.

\vskip 5mm

{\bf Fig.5.} The fractional population oscillations for $b=0.4$
and $\delta=0.2$.

\vskip 5mm

{\bf Fig.6.} The critical population dynamics for $b=0.499$
and $\delta=0.001001002$. As in all previous figures, solid line corresponds
to $n_j$ and the dashed one to $n_0$.


\newpage

\begin{thebibliography}{99}

\bibitem{1}
Parkins, A.C. and Walls, D.F., 1998, {\it Phys. Rep.}, {\bf 303}, 1.

\bibitem{2}
Dalfovo, F., Giorgini, S., Pitaevskii, L.P., and Stringari, S., 1999,
{\it Rev. Mod. Phys.}, {\bf 71}, 463.

\bibitem{3}
Yukalov, V.I., Yukalova E.P., and Bagnato, V.S., 1997, {\it Phys. Rev. A},
{\bf 56}, 4845.

\bibitem{4}
Yukalov, V.I., 1998, {\it Statistical Green's Functions} (Kingston:
Queen's University).

\bibitem{5}
Gross, E.P., 1957, {\it Phys. Rev.}, {\bf 106}, 161.

\bibitem{6}
Ginzburg, V.L. and Pitaevskii, L.P., 1958, {\it J. Exp. Theor. Phys.},
{\bf 7}, 858.

\bibitem{7}
Pitaevskii, L.P., 1961, {\it J. Exp. Theor. Phys.}, {\bf 13}, 451.

\bibitem{8}
Bogolubov, N.N. and Mitropolsky, Y.A., 1961, {\it Asymptotic Methods in the
Theory of Nonlinear Oscillations} (New York: Gordon and Breach).

\bibitem{9}
Weiner, J., Bagnato, V.S., Zilio, S.C., and Julienne, P., 1999, {\it Rev.
Mod. Phys.}, {\bf 71}, 1.

\bibitem{10}
Williams, J., Walser R., Cooper, J., Cornell, E.A., and Holland, M., 1999,
preprint cond-mat/9904399.

\bibitem{11}
Mewes, M.O. et al., 1997, {\it Phys. Rev. Lett.}, {\bf 78}, 582.

\bibitem{12}
Andrews, M.R. et al., 1997, {\it Science}, {\bf 275}, 637.

\bibitem{13}
Burt, E.A. et al., 1997, {\it Phys. Rev. Lett.}, {\bf 79}, 337.

\bibitem{14}
Bloch, I., H\"ansch, T.W., and Esslinger, T., 1999, {\it Phys. Rev. Lett},
{\bf 82}, 3008.

\bibitem{15}
Hagley, E.W. et al., 1999, {\it Science}, {\bf 283}, 1706.


\end{thebibliography}
\end{document}